\documentclass[conference]{IEEEtran}
\IEEEoverridecommandlockouts
\usepackage{cite}
\usepackage{amsmath,amssymb,amsfonts}
\usepackage{algorithmic}
\usepackage{siunitx}
\usepackage{graphicx}
\usepackage{textcomp}
\usepackage{xcolor}
\usepackage[a4paper, total={184mm,239mm}]{geometry}
\def\BibTeX{{\rm B\kern-.05em{\sc i\kern-.025em b}\kern-.08em
        T\kern-.1667em\lower.7ex\hbox{E}\kern-.125emX}}

\usepackage{fancyhdr}
\usepackage[all]{background}
\usepackage{stackengine}
\setstackEOL{\\}
\setstackgap{L}{\normalbaselineskip}
\SetBgContents{\color{gray}{\tiny \Longstack{PREPRINT - accepted In Proceedings of the 60th Design Automation Conference  (DAC '23)}}}% Set contents
\SetBgPosition{4.2cm,1cm}% Select location
\SetBgOpacity{1.0}% Select opacity
\SetBgAngle{0}% Select rotation of logo
\SetBgScale{1.8}% Select scale factor of logo
\usepackage{pgfplots}
\usepackage{pgfplotstable}
\pgfplotsset{compat=1.16} % use 1.17 for apple, and 1.16 for ubuntu!

\usetikzlibrary{patterns, arrows, pgfplots.groupplots,hobby}
\usetikzlibrary{patterns}
\usetikzlibrary{positioning}
\usetikzlibrary{pgfplots.groupplots}
\usetikzlibrary{plotmarks}
\usetikzlibrary{calc}
\usepgfplotslibrary{external}
\usepackage{balance}
\usepackage{caption}
\usepackage{xcolor}
\usepackage{colortbl}
\usepackage{rotating}
\usepackage{multirow}
\usepackage{booktabs}
\usepackage{bm}
\usepackage{subfig}
\usepackage{enumitem}
\usepackage{soul}
\usepackage[export]{adjustbox}
\usepackage{bigstrut}
\usepackage{graphicx}
\usepackage{amsmath}
\definecolor{cg1}{HTML}{E8F1FA}
\definecolor{cg2}{HTML}{C7DDF2}
\definecolor{cg3}{HTML}{8EBAE5}
\definecolor{cg4}{HTML}{407FB7}
\definecolor{cg5}{HTML}{00549F}
\definecolor{red}{HTML}{CC071E}

\usepackage{balance}

\definecolor{cg1}{HTML}{9C6261}
\definecolor{cg2}{HTML}{E98667}
\definecolor{cg3}{HTML}{E8B78E}
\definecolor{cg4}{HTML}{DE9853}
\definecolor{cg5}{HTML}{775445}
\definecolor{cg6}{HTML}{3C3D28}
\usepackage{keyval}
\usepackage{epstopdf}
\usepackage{upgreek}
\usepackage{float}

\begin{document}
\bstctlcite{IEEEexample:BSTcontrol}
\title{Fault Injection in Native Logic-in-Memory Computation on Neuromorphic Hardware
\thanks{This work was funded by the Federal Ministry of Education and Research(BMBF, Germany) in the project NEUROTEC II (16ME0398K, 16ME0399).}
%\vspace{-4mm}
}
\author{
    \IEEEauthorblockN{Felix Staudigl\IEEEauthorrefmark{1},
                      Thorben Fetz\IEEEauthorrefmark{1},
                      Rebecca Pelke\IEEEauthorrefmark{1},
                      Dominik Sisejkovic\IEEEauthorrefmark{1},\\
                      Jan Moritz Joseph\IEEEauthorrefmark{1},
                      Leticia Bolzani P\"ohls\IEEEauthorrefmark{2}, and
                      Rainer Leupers\IEEEauthorrefmark{1}
    }
    \IEEEauthorblockA{\IEEEauthorrefmark{1}
        \textit{Institute for Communication Technologies and Embedded Systems, RWTH Aachen University, Germany}
    }
    \IEEEauthorblockA{\IEEEauthorrefmark{2}
        \textit{Chair of Integrated Digital Systems and Circuit Design, RWTH Aachen University, Germany}\\
    \{staudigl, fetz, pelke, sisejkovic, joseph, leupers\}@ice.rwth-aachen.de\\
    poehls@ids.rwth-aachen.de\\
    }
\vspace{-9mm}
%\vspace{20mm}
}

\maketitle

\begin{abstract}
Logic-in-memory (LIM) describes the execution of logic gates within memristive crossbar structures, promising to improve performance and energy efficiency. Utilizing only binary values, LIM particularly excels in accelerating binary neural networks, shifting it in the focus of edge applications. Considering its potential, the impact of faults on BNNs accelerated with LIM still lacks investigation. In this paper, we propose faulty logic-in-memory (\textbf{FLIM}), a fault injection platform capable of executing full-fledged BNNs on LIM while injecting in-field faults. The results show that FLIM runs a single MNIST picture 66754$\times$ faster than the state of the art by offering a fine-grained fault injection methodology.
\end{abstract}

\begin{IEEEkeywords}
ReRAM, memristor, faults, reliability, logic-in-memory\vspace{-3mm}
\end{IEEEkeywords}

\section{Introduction}
The von Neumann architecture describes a computing system consisting of two main distinct components: the memory and the computing unit. The computing unit must fetch/push from/to the memory in order to process data, representing the so-called von Neumann bottleneck. The bottleneck drastically limits conventional computing systems' performance and energy efficiency. Consequently, novel computing paradigms are being investigated to overcome this limitation~\cite{Staudigl2022a}.

\begin{figure}[t]
    \centering
    \includegraphics[width=0.8\columnwidth]{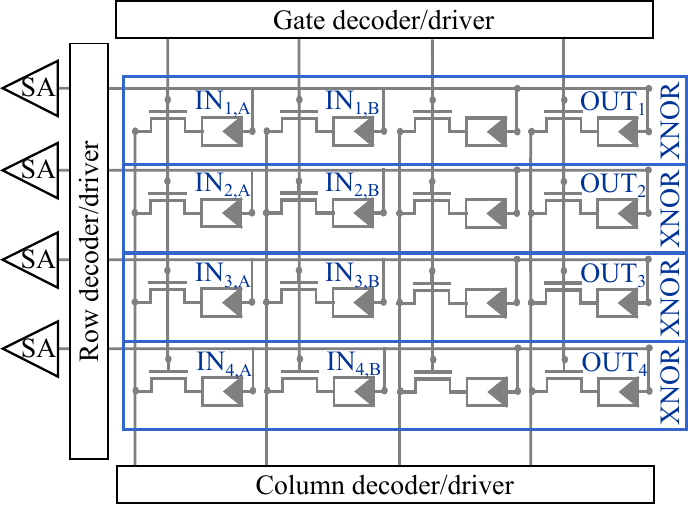}
    \caption{Memristive crossbar array executing parallel XNOR operations.}
    \label{fig:intro_lim}
    \vspace{-5mm}
\end{figure}

Emerging non-volatile memories such as spin-torque-transfer memory (STT-RAM/MRAM), phase-change random-access memory (PCRAM), and resistive random-access memory (ReRAM) provide an ideal substrate for high-density memories by also enabling the computing-in-memory (CIM) paradigm. CIM executes operations within the memory without moving data to the processing unit.
Implementing these operations in an analog fashion requires expensive ADCs/DACs but accomplishes the best performance~\cite{Ankit2019}. In comparison, logic-in-memory (LIM) uses binary values to perform logic operations within memory, omitting the conversion from the analog to the digital domain, while being more resilient against technology-specific non-idealities~\cite{Gaillardon2016,Papandroulidakis2017}. Fig.~\ref{fig:intro_lim} exemplifies a memristive crossbar array executing parallel XNOR operations.

Binary neural networks (BNNs) represent a set of machine learning models that replace the typically used full-precision weights with binary values. These networks trade a lower overall accuracy with a significant performance improvement and a lower memory footprint. Due to the quantification of its internal layers, the inference is dominantly computed through the XNOR operation~\cite{Qin2020a}. Hence, BNNs benefit from the massive parallelization of LIM, particularly in the context of edge applications.

\begin{figure*}[ht!]
    \centering
    \includegraphics[width=\textwidth]{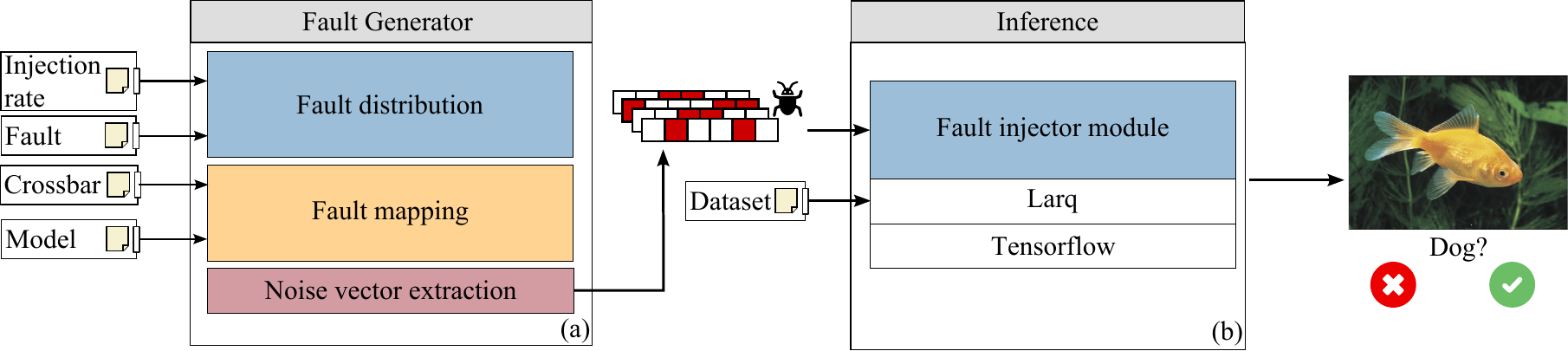}
    \caption{Overview of the simulation methodology: (a) Noise vector generator and (b) fault injector.}
    \label{fig:sim_overview}
    \vspace{-5mm}
\end{figure*}
However, the benefit of non-volatile memories for implementing these emerging applications depends on being able to guarantee reliability during their lifetime. In more detail, as observed in CMOS-based memories, these novel memories are susceptible to time-dependent deviations, causing in-field faults that affect their lifetime reliability~\cite{Vatajelu2018,Li2019}. Time-dependent deviations are primarily a result of environmental variations, causing transient faults, such as bit-flips, and temporal variations, causing degradation over a lifetime. Furthermore, towards the end of their life cycle, memories encounter stuck-at faults.

The impact of transient faults has been thoroughly investigated for analog CIM~\cite{Rasch2021}. Unfortunately, there is only limited work on the effect on LIM. X-Fault~\cite{Staudigl2022} describes the most detailed end-to-end fault injection platform injecting different traditional faults at the device level. However, this approach limits the platform's performance, dramatically lowering the feasibility of real-world models and datasets.

\textbf{Contributions:}~In this paper, we propose an ultra-fast fault injection platform called FLIM, capable of simulating full-fledged BNN models. FLIM processes an MNIST data frame 66754$\times$ faster than X-Fault while injecting different faults related to time-dependent deviations. In detail, we present the following investigations. (1) First, we develop a simulation methodology that abstracts in-field faults toward a high-performance fault model. (2) Second, we introduce a notion of time within our simulator, which allows the injection of faults per layer. (3) Finally, we perform a reliability assessment considering in-field faults of BNNs using different datasets and models.

The rest of the paper is organized as follows. Section~\ref{sec:back} summarizes the state-of-the-art fault injection platforms, as well as the background related to BNNs and LIM. Section~\ref{sec:methodology}, describes the simulation methodology, including the implemented fault models. Experimental results and a detailed discussion are presented in Section~\ref{sec:res}. Section~\ref{sec:con} concludes the paper.

\section{Background}
\label{sec:back}
\begin{figure*}[ht!]
    \centering
    \includegraphics[width=\textwidth]{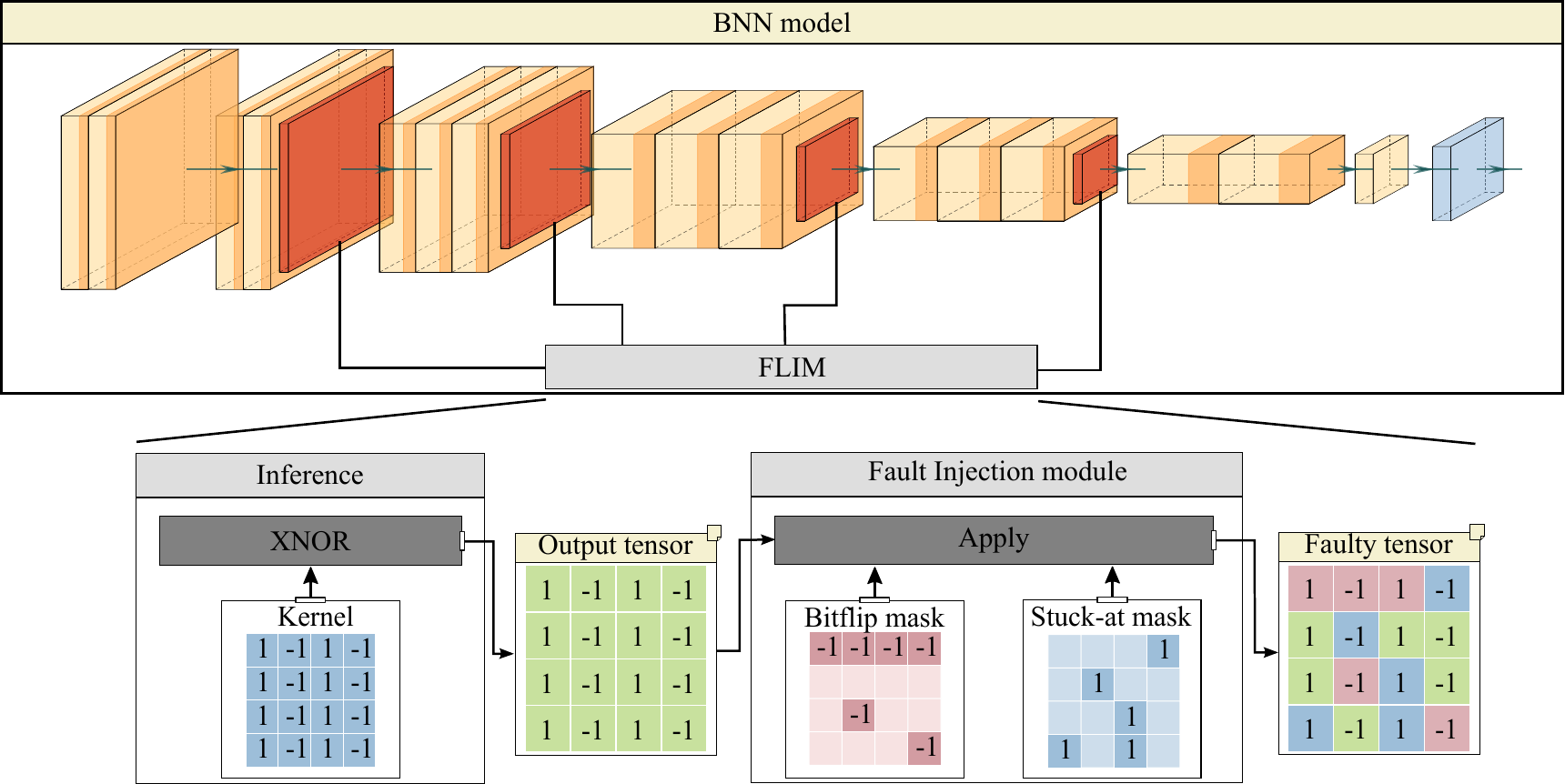}
    \caption{Internal structure of FLIM consisting of the Fault Generator and the Fault Injector module.}
    \label{fig:sim_internals}
    \vspace{-5mm}
\end{figure*}
This section summarizes the main approaches proposed for performing reliability and security assessments, considering different types of faults that can affect these novel applications after manufacturing and during their lifetime. In addition, it describes the background of BNNs and LIM.

\subsection{Related Work}
Fault injection platforms allow for a hands-on investigation of the impact of faults on various aspects of computing systems. For instance, these platforms are heavily used to investigate hardware security primitives~\cite{Gay2019,Santos2021,Wang2022}. Especially analyzing faults in machine learning algorithms has become a vital field of research considering their widespread usage~\cite{GuerreroBalaguera2022,Spyrou2021,Liu2017}. Non-idealities of emerging non-volatile memories and their impact on CIM have been thoroughly investigated~\cite{Kannan2015,Chen2015}. Chakraborty et al.~\cite{Chakraborty2020} propose a general approach to model faults on memristive crossbars executing neural networks. The framework is capable of simulating linear and non-linear non-idealities at an architectural level. PytorX~\cite{He2019} presents an end-to-end neural network tool based on PyTorch. The tool adjusts the mapping and optimizes the training to overcome the effect of non-ideal crossbars, and drastically limits the impact of faults. In general, existing research has mainly focused on analog-based CIM. The only framework able to simulate LIM on memristive crossbar is presented by X-Fault~\cite{Staudigl2022}. The framework offers a wide range of features, including various fault models and injection mechanisms. However, the tool simulates faults on memristor level limiting performance significantly. Hence, X-Fault cannot simulate larger models or datasets due to performance issues. Consequently, we propose FLIM, which closes this gap and allows for an extensive investigation of faults on LIM-based machine learning algorithms.

\subsection{Binary Neural Networks (BNNs)}
BNNs represent a class of neural networks using aggressive quantization, drastically improving power efficiency but reducing accuracy~\cite{Qin2020}. This approach is auspicious for deploying deep neural networks to resource-constrained devices. Compared to full-precision neural networks, BNNs are behind in terms of accuracy. However, simple classification tasks achieve competitive performance.  The open-source library Larq~\cite{Geiger2020} offers an easy entry to build and train BNNs. The library builds upon Tensorflow and provides pre-trained models. An XNOR operation within BNNs replaces the matrix-matrix-multiplication of convolutions in full-precision neural networks.
Thus, BNNs map directly to LIM on memristive crossbars and position it as a preferred application for CIM on edge. Since BNNs still require some non-binary computation (e.g., activation and integer bit-count), only convolutional and dense layers are mapped on memristive crossbar arrays. We follow X-Fault's conservative approach by assuming that these non-binary operations are executed in CMOS.

\subsection{Logic-in-Memory (LIM)}
Compared to conventional CIM, LIM utilizes the memristive crossbar array in a binary fashion omitting expensive ADCs/DACS. Due to its binary working mode, LIM trades a higher error resilience with lower latency. Internally, logical states (0 or 1) are represented as either high or low resistive values of the memristive cell. Logic gates are composed of multiple memristors. An operation voltage applied to the connecting word line calculates the respective output based on the given inputs. Kvatinsky et al.~\cite{Kvatinsky2014} classified logic families into three categories: statefulness, the proximity of computation, and flexibility. MAGIC~\cite{Kvatinsky2014} and IMPLY~\cite{Kvatinsky2014a} describe two stateful logic families capable of implementing a complete set of logic operations.
Within the scope of this work, we abstracted the computation to the application level. Hence, we assume the underlying usage of a logic family implementing the XNOR logic gate without modeling it in detail.

\section{Faulty Logic-in-Memory (FLIM)}
\label{sec:methodology}
FLIM embodies an end-to-end simulator capable of emulating the impact of faults on BNNs using binary memristive crossbar arrays. Fig.~\ref{fig:sim_overview} depicts the internal structure of the fault injection platform, which consists of a \textit{Fault Generator} and a \textit{Fault Injector}. The \textit{Fault Generator} constructs a set of fault vectors encoding the fault type, location, and injection rate. This tool is implemented in vanilla Python and hence, independent of the fault injection mechanism. The \textit{Fault Injector} extends the Tensorflow/Larq framework to dynamically inject faults in arbitrary BNN models. As depicted in Fig.~\ref{fig:sim_overview}b, the \textit{Fault Injector} employs the previously generated fault vectors and a defined dataset to initiate the inference procedure.

\begin{figure*}[ht!]
    \centering
    \subfloat[]{
        \begin{tikzpicture}
        % bitlfip layers
            \pgfplotstableread[row sep=\\, col sep=&]{
            bitflip & conv2d1 & conv2d2 & dense0 & dense1 & combined\\
            0.0 & 98.37 & 98.37 & 98.37 & 98.37 & 98.37 \\
            5.0 & 97.96 & 97.97 & 98.2 & 98.18 & 97.41 \\
            10.0 & 96.94 & 97.8 & 97.74 & 97.93 & 94.05 \\
            15.0 & 93.52 & 96.65 & 97.63 & 97.68 & 77.2 \\
            20.0 & 82.25 & 95.19 & 96.69 & 96.44 & 47.82 \\
            25.0 & 72.16 & 88.14 & 94.37 & 92.88 & 23.02 \\
            30.0 & 23.33 & 73.86 & 88.23 & 81.45 & 15.05\\
            }\dataset
            \begin{axis}[
                name=plot1,
                ylabel={Accuracy (\%)},
                xlabel={Injection rate (\%)},
                ylabel style={at={(-0.23, 0.5)}, anchor=north},
                legend style={at={(0.25,0.6)}, anchor=north, legend cell align={left}, nodes={scale=0.85, transform shape}},
                legend image post style={scale=1},
                legend columns=1,
                width=0.33\textwidth,
                ymajorgrids=true,
                xmajorgrids=true,
                height=5cm,
                ymin=0,
                ymax=110,
                ytick={0, 10, 20, 30, 40, 50, 60, 70, 80, 90, 100},
                xtick={0, 5, 10, 15, 20, 25, 30},
                ylabel style ={font=\small},
                xlabel style ={font=\small},
                tick label style={font=\small}
                ]
                \addplot[draw=cg1, thick] table[x=bitflip, y=conv2d1]{\dataset};
                \addplot[draw=cg2, thick, densely dotted] table[x=bitflip, y=conv2d2]{\dataset};
                \addplot[draw=cg3, thick, densely dashed] table[x=bitflip, y=dense0]{\dataset};
                \addplot[draw=cg4, thick, dashdotted] table[x=bitflip, y=dense1]{\dataset};
                \addplot[draw=red, thick, mark=square*,mark options={fill=red}] table[x=bitflip, y=combined]{\dataset};
                \legend{conv\_1, conv\_2, dense\_0, dense\_1, combined}
            \end{axis}
        \end{tikzpicture}
    }\hfil
    \subfloat[]{
        \begin{tikzpicture}
        % stuck-at layers
            \pgfplotstableread[row sep=\\, col sep=&]{
            stuckat & conv2d1 & conv2d2 & dense0 & dense1 & combined\\
            0.0 & 98.37 & 98.37 & 98.37 & 98.37 & 98.37 \\
            5.0 & 93.78 & 92.21 & 96.11 & 98.32 & 78.96 \\
            10.0 & 52.86 & 27.82 & 52.5 & 98.27 & 12.34 \\
            15.0 & 11.81 & 9.32 & 10.5 & 97.94 & 9.79 \\
            20.0 & 11.35 & 8.92 & 9.36 & 97.85 & 10.04 \\
            25.0 & 11.35 & 8.92 & 9.58 & 97.64 & 10.38 \\
            30.0 & 11.35 & 8.92 & 9.58 & 97.9 & 10.67\\
            }\dataset
            \begin{axis}[
                ylabel={Accuracy (\%)},
                xlabel={Injection rate (\%)},
                ylabel style={at={(-0.23, 0.5)}, anchor=north},
                legend style={at={(0.73,0.8)}, anchor=north, nodes={scale=0.85, transform shape}},
                legend image post style={scale=1.2},
                legend columns=1,
                width=0.33\textwidth,
                ymajorgrids = true,
                xmajorgrids = true,
                height=5cm,
                ymin=0,
                ymax=110,
                ytick={10, 20, 30, 40, 50, 60, 70, 80, 90, 100},
                xtick={0, 5, 10, 15, 20, 25, 30},
                ylabel style ={font=\small},
                xlabel style ={font=\small},
                tick label style={font=\small}
                ]
                \addplot[draw=cg1, thick] table[x=stuckat, y=conv2d1]{\dataset};
                \addplot[draw=cg2, thick, densely dotted] table[x=stuckat, y=conv2d2]{\dataset};
                \addplot[draw=cg3, thick, densely dashed] table[x=stuckat, y=dense0]{\dataset};
                \addplot[draw=cg4, thick, dashdotted] table[x=stuckat, y=dense1]{\dataset};
                \addplot[draw=red, thick, mark=square*,mark options={fill=red}] table[x=stuckat, y=combined]{\dataset};
            \end{axis}
        \end{tikzpicture}
    }\hfil
    \subfloat[]{
        \begin{tikzpicture}
        % dynamic layers
            \pgfplotstableread[row sep=\\, col sep=&]{
            dynamic & conv2d1 & conv2d2 & dense0 & dense1 & combined\\
            0.0 & 26.08 & 73.11 & 88.02 & 85.81 & 15.52 \\
            1.0 & 93.42 & 96.49 & 97.37 & 89.36 & 68.54 \\
            2.0 & 96.46 & 97.37 & 97.81 & 96.24 & 84.49 \\
            3.0 & 97.26 & 97.91 & 97.98 & 95.06 & 90.43 \\
            4.0 & 97.7 & 97.97 & 97.95 & 96.66 & 93.68\\
            }\dataset
            \begin{axis}[
                ylabel={Accuracy (\%)},
                xlabel={\# of XNOR ops},
                ylabel style={at={(-0.23, 0.5)}, anchor=north},
                legend style={at={(0.5,1.25)}, anchor=north, legend cell align={left}},
                legend image post style={scale=0.7},
                legend columns=3,
                width=0.33\textwidth,
                ymajorgrids=true,
                xmajorgrids=true,
                height=5cm,
                ymin=0,
                ymax=110,
                ytick={10, 20, 30, 40, 50, 60, 70, 80, 90, 100},
                xtick={0, 1, 2, 3, 4},
                ylabel style ={font=\small},
                xlabel style ={font=\small},
                tick label style={font=\small}
                ]
                \addplot[draw=cg1, thick] table[x=dynamic, y=conv2d1]{\dataset};
                \addplot[draw=cg2, thick, densely dotted] table[x=dynamic, y=conv2d2]{\dataset};
                \addplot[draw=cg3, thick, densely dashed] table[x=dynamic, y=dense0]{\dataset};
                \addplot[draw=cg4, thick, dashdotted] table[x=dynamic, y=dense1]{\dataset};
                \addplot[draw=red, thick, mark=square*,mark options={fill=red}] table[x=dynamic, y=combined]{\dataset};
            \end{axis}
        \end{tikzpicture}
    }

    \subfloat[]{
        \begin{tikzpicture}
        % faulty column
            \pgfplotstableread[row sep=\\, col sep=&]{
            cols & conv2d1 & conv2d2 & dense0 & dense1 & combined\\
            0.0 & 98.37 & 98.37 & 98.37 & 98.37 & 98.37 \\
            1.0 & 96.82 & 97.77 & 97.7 & 97.97 & 93.06 \\
            2.0 & 83.45 & 93.57 & 96.87 & 96.96 & 52.59 \\
            3.0 & 33.21 & 73.28 & 88.3 & 87.77 & 13.12 \\
            4.0 & 10.93 & 37.22 & 52.16 & 37.29 & 9.56\\
            }\dataset
            \begin{axis}[
                ylabel={Accuracy (\%)},
                xlabel={\# of faulty columns},
                ylabel style={at={(-0.23, 0.5)}, anchor=north},
                legend style={at={((0.25,0.6)}, anchor=north, legend cell align={left}, nodes={scale=0.85, transform shape}},
                width=0.33\textwidth,
                legend image post style={scale=1},
                ymajorgrids=true,
                xmajorgrids=true,
                height=5cm,
                ymin=0,
                ymax=110,
                ytick={10, 20, 30, 40, 50, 60, 70, 80, 90, 100},
                xtick={0, 1, 2, 3, 4},
                ylabel style ={font=\small},
                xlabel style ={font=\small},
                tick label style={font=\small}
                ]
                \addplot[draw=cg1, thick] table[x=cols, y=conv2d1]{\dataset};
                \addplot[draw=cg2, thick, densely dotted] table[x=cols, y=conv2d2]{\dataset};
                \addplot[draw=cg3, thick, densely dashed] table[x=cols, y=dense0]{\dataset};
                \addplot[draw=cg4, thick, dashdotted] table[x=cols, y=dense1]{\dataset};
                \addplot[draw=red, thick, mark=square*,mark options={fill=red}] table[x=cols, y=combined]{\dataset};
            \end{axis}
        \end{tikzpicture}
    }\hfill
    \subfloat[]{
        \begin{tikzpicture}
        % faulty column
            \pgfplotstableread[row sep=\\, col sep=&]{
            rows & conv2d1 & conv2d2 & dense0 & dense1 & combined\\
            0.0 & 98.37 & 98.37 & 98.37 & 98.37 & 98.37 \\
            1.0 & 98.21 & 98.01 & 98.23 & 98.26 & 97.84 \\
            2.0 & 97.48 & 97.92 & 98.23 & 97.88 & 95.73 \\
            3.0 & 97.06 & 97.85 & 98.06 & 92.25 & 87.11 \\
            4.0 & 95.49 & 97.55 & 97.99 & 87.09 & 78.69 \\
            5.0 & 91.89 & 96.93 & 97.77 & 84.39 & 69.1 \\
            6.0 & 87.4 & 96.06 & 97.44 & 81.87 & 57.78 \\
            7.0 & 80.36 & 94.32 & 96.99 & 78.54 & 46.6 \\
            8.0 & 70.51 & 91.24 & 96.43 & 76.42 & 35.68 \\
            9.0 & 60.61 & 86.72 & 95.45 & 76.29 & 27.59 \\
            10.0 & 47.55 & 81.36 & 93.87 & 72.47 & 19.27 \\
            11.0 & 37.49 & 74.28 & 91.53 & 68.67 & 16.24 \\
            12.0 & 29.09 & 65.54 & 88.08 & 66.04 & 13.52 \\
            13.0 & 22.15 & 57.41 & 82.63 & 64.0 & 11.47 \\
            14.0 & 17.8 & 46.86 & 74.24 & 60.35 & 10.33 \\
            15.0 & 14.99 & 39.38 & 63.14 & 59.17 & 9.84 \\
            16.0 & 12.65 & 31.2 & 51.82 & 55.93 & 9.56 \\
            17.0 & 11.95 & 24.17 & 38.21 & 54.16 & 9.89 \\
            18.0 & 11.27 & 18.55 & 27.11 & 51.31 & 10.07 \\
            19.0 & 10.75 & 14.03 & 16.65 & 47.45 & 9.82\\
            }\dataset
            \begin{axis}[
                ylabel={Accuracy (\%)},
                xlabel={\# of affacted rows},
                ylabel style={at={(-0.23, 0.5)}, anchor=north},
                legend style={at={(0.5,1.25)}, anchor=north, legend cell align={left}},
                legend image post style={scale=0.7},
                legend columns=3,
                ymajorgrids=true,
                xmajorgrids=true,
                width=0.33\textwidth,
                height=5cm,
                ymin=0,
                ymax=110,
                ytick={10, 20, 30, 40, 50, 60, 70, 80, 90, 100},
                xtick={0, 2, 4, 6, 8, 10, 12, 14, 16, 18, 20},
                ylabel style ={font=\small},
                xlabel style ={font=\small},
                tick label style={font=\small}
                ]
                \addplot[draw=cg1, thick] table[x=rows, y=conv2d1]{\dataset};
                \addplot[draw=cg2, thick, densely dotted] table[x=rows, y=conv2d2]{\dataset};
                \addplot[draw=cg3, thick, densely dashed] table[x=rows, y=dense0]{\dataset};
                \addplot[draw=cg4, thick, dashdotted] table[x=rows, y=dense1]{\dataset};
                \addplot[draw=red, thick, mark=square*,mark options={fill=red}] table[x=rows, y=combined]{\dataset};
            \end{axis}
        \end{tikzpicture}
    }\hfill
    \subfloat[]{
        % performance comparison
        \begin{tikzpicture}[scale=0.455]
            \pgfplotstableread[row sep=\\,col sep=&]{
                plat & CPU & GPU \\
                0  &   1882000 & 1786000 \\
                25 &   64.067  & 26.755 \\
                50 &   2.29    & 2.603 \\
            }\dataset
            \begin{axis}[ybar,
                ylabel={Runtime in \SI{}{\second}},
                xlabel={},
                ylabel style={at={(-0.15,0.5)},anchor=north},
                legend style={at={(0.5,0.96), font=\Large},
                    anchor=north,legend cell align={left}},
                legend image post style={scale=1.4},
                enlarge x limits={abs=1.2cm},
                ymajorgrids = true,
                bar width=.5cm,
                 ymin = 0,
                ymode = log,
                width=12cm,
                height=10cm,
                ymax=10000000,
                ytick={1, 10, 100, 1000, 10000, 100000, 1000000, 10000000},
                xtick align=inside,
                xtick=data,
                xticklabel style={align=center},
                xticklabels = {
                    X-Fault,
                    FLIM,
                    Vanilla Larq
                },
                ylabel style ={font=\huge},
                xlabel style ={font=\huge},
                tick label style={font=\Large},
                legend entries={CPU, CPU+GPU},
                legend columns=2,
                ]
                \addplot[draw=black,fill=white] table[x=plat,y=CPU] \dataset;
                \addplot[draw=black,fill=cg4]   table[x=plat,y=GPU] \dataset;
            \end{axis}
        \end{tikzpicture}
    }
    \caption{Simulation results: Impact of (a) bit-flips, (b) stuck-at, (c) dynamic faults, (d) faulty columns, and (e) faulty rows on different layers. (f) Performance benchmark.}
    \label{fig:fault_layers}
\vspace{-4mm}
\end{figure*}
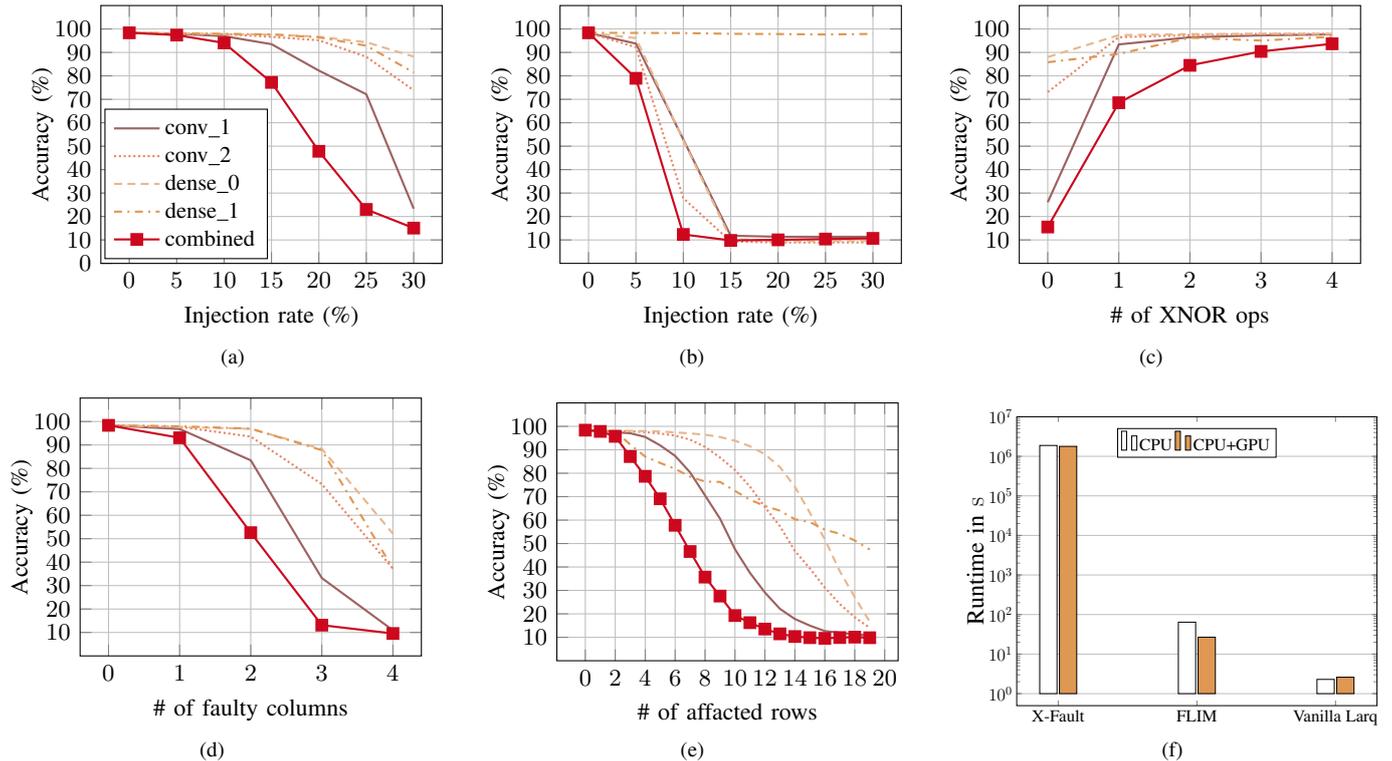

\textbf{Fault masking:} FLIM implements bit-flip and stuck-at faults to investigate the impact of time-dependent deviations. In contrast to X-Fault, the proposed platform models and injects faults on the XNOR operation level yielding enhanced simulation performance.

For bit-flip and stuck-at faults, a \textit{fault mask} is generated, encoding the fault's location and binary representation. The bit-flip mask defines a 2-dimensional Boolean array initialized with zeros. The injection rate specifies the number of elements within the array set to 1. In addition to these randomly distributed bit-flips, entire rows/columns may also be faulty. Thus, these rows/columns are set to 1, respectively.

Furthermore, the platform supports \textit{dynamic faults} which occur every n-th XNOR operation~\cite{Priya2014}. To model dynamic faults, the fault mask has to be repeated over several layers. Therefore, multiple bit-flip masks are assembled, which are consecutively applied to the respective layers of the model during inference. Likewise, the stuck-at mask follows the same structure by initializing a 2-dimensional array with zeros and marking all faulty elements with ones. In general, mask generation happens as an offline process that significantly improves performance because the expensive mapping and distribution of faults are performed once and reused over the whole simulation.

\textbf{Fault mapping:} The generated masks are assigned to specific layers within the BNN model in the next step. Therefore, the \textit{Fault Generator} has to be provided with the dimensions and the number of crossbars used during the simulation. First, the mapping tool calculates the number of parallel XNOR operations based on the crossbars. Considering the implementation of MAGIC~\cite{Kvatinsky2014} or IMPLY~\cite{Kvatinsky2014a}, four memristors are required to facilitate one XNOR operation. Second, the model extracts the total number of required XNOR operations. Within a BNN, only the 2-dimensional convolution (conv2D) layer and fully binarized dense layers are dominantly using the XNOR operation. Consequently, these layers would be mapped and accelerated onto memristive crossbar arrays, while all remaining layers are executed on conventional CMOS. Hence, the mapping tool extracts the dimensions of these layers and assigns the previously generated fault masks.

\textbf{Fault vector extraction:} Finally, the required fault vectors are extracted from the virtual crossbar representation. The 2-dimensional arrays are flattened to 1 dimension. Furthermore, the vectors are stored in a binary file annotated with meta-information about the assigned layer and mask type. The binary file is independent of the dataset and reusable for a myriad of experiments.

\textbf{Fault Injector:} The \textit{Fault Injector} represents the centerpiece of the FLIM platform. The tool is deeply integrated with the Larq and Tensorflow framework to aim for maximum performance by achieving a granular fault injection.

Fig.~\ref{fig:sim_internals} depicts the internals of the injection mechanism. The Larq library generally extends the Keras framework to facilitate BNNs~\cite{Geiger2020}. Larq defines custom quantized layers as an extension of Keras layers. We extended this layer base class by adding an instance of the Fault Injector. To trigger the injection mechanism during the inference, the original convolution method has been overwritten. The following describes the procedure of the faulty convolution method. First, the standard convolution function calculates the feature map. The feature map does not yet take into account any faults and, therefore, represents the correct result of the computation. Second, before both fault masks are applied on the feature map, the vectors must be adjusted in length depending on the batch size and the input dimension. Finally, the fault masks are applied by performing another XNOR operation.

\section{Results and Discussion}
\label{sec:res}
\begin{figure*}[ht]
    \centering
    \subfloat[]{
        \begin{tikzpicture}
        % bitlfip zoo
            \pgfplotstableread[row sep=\\, col sep=&]{
            bitflip & BinaryDenseNet37Dilated & BinaryDenseNet45 & BinaryDenseNet37 & BinaryDenseNet28 & BinaryResNetE18 & RealToBinaryNet & BinaryAlexNet & MeliusNet22 & BiRealNet & XNORNet\\
            0.0 & 45.54 & 58.2 & 63.04 & 36.26 & 65.78 & 58.66 & 61.46 & 63.66 & 64.02 & 36.32 \\
            1.0 & 43.94 & 55.26 & 62.54 & 35.98 & 64.82 & 57.02 & 61.74 & 63.1 & 63.64 & 37.26 \\
            2.0 & 41.36 & 50.52 & 62.18 & 32.66 & 64.92 & 54.2 & 60.5 & 62.48 & 63.42 & 36.98 \\
            3.0 & 40.44 & 43.24 & 59.94 & 29.06 & 64.36 & 50.44 & 59.28 & 60.38 & 62.32 & 35.5 \\
            4.0 & 39.12 & 31.44 & 57.72 & 27.0 & 63.8 & 48.56 & 56.78 & 58.8 & 59.78 & 35.12 \\
            5.0 & 35.36 & 15.64 & 54.42 & 20.76 & 63.04 & 43.22 & 56.08 & 57.58 & 56.48 & 31.56 \\
            6.0 & 26.86 & 12.98 & 51.44 & 12.5 & 61.8 & 26.16 & 52.92 & 52.2 & 52.9 & 28.2 \\
            7.0 & 23.7 & 5.08 & 45.2 & 10.34 & 58.36 & 25.54 & 50.5 & 48.38 & 47.08 & 23.52 \\
            8.0 & 21.94 & 3.68 & 41.16 & 3.36 & 56.74 & 12.42 & 44.88 & 44.36 & 44.78 & 19.96 \\
            9.0 & 13.74 & 2.62 & 27.32 & 2.5 & 54.98 & 7.18 & 43.12 & 41.14 & 35.98 & 15.7 \\
            10.0 & 9.28 & 1.62 & 23.6 & 1.3 & 51.56 & 5.68 & 32.96 & 30.62 & 25.24 & 13.38 \\
            11.0 & 5.86 & 0.64 & 16.72 & 0.68 & 48.6 & 2.28 & 33.22 & 26.6 & 20.34 & 9.32 \\
            12.0 & 2.84 & 0.28 & 10.14 & 0.34 & 43.58 & 1.58 & 26.4 & 17.74 & 15.42 & 7.78 \\
            13.0 & 2.26 & 0.46 & 4.38 & 0.3 & 36.66 & 1.44 & 17.1 & 12.02 & 10.06 & 7.64 \\
            14.0 & 1.24 & 0.28 & 3.38 & 0.16 & 33.42 & 0.9 & 16.38 & 11.72 & 6.54 & 5.16 \\
            15.0 & 1.38 & 0.34 & 3.02 & 0.1 & 29.58 & 0.6 & 10.94 & 7.08 & 4.64 & 3.74 \\
            16.0 & 0.42 & 0.22 & 1.32 & 0.14 & 25.4 & 0.54 & 6.66 & 4.16 & 4.04 & 2.5 \\
            17.0 & 0.34 & 0.16 & 2.24 & 0.1 & 17.04 & 0.46 & 5.56 & 4.64 & 4.18 & 2.86 \\
            18.0 & 0.22 & 0.26 & 1.42 & 0.12 & 14.48 & 0.38 & 4.92 & 3.58 & 3.12 & 2.48 \\
            19.0 & 0.2 & 0.06 & 0.94 & 0.14 & 8.42 & 0.48 & 3.22 & 2.62 & 1.9 & 2.36 \\
            20.0 & 0.18 & 0.1 & 0.84 & 0.08 & 5.22 & 0.4 & 2.76 & 2.66 & 1.88 & 1.64\\
            }\dataset
            \begin{axis}[
                name=plot1,
                ylabel={Accuracy (\%)},
                xlabel={Injection rate (\%)},
              ylabel style={at={(-0.06, 0.5)}, anchor=north},
             legend style={at={(0.5,1.5)}, anchor=north, legend cell align={left}, nodes={scale=0.85, transform shape}},
             legend image post style={scale=1},
                legend columns=5,
                width=\textwidth,
                ymajorgrids=true,
                xmajorgrids=true,
                height=3.5cm,
                ymin=-5, ymax=70,
                xmin=-.5, xmax=20.5,
                ytick={0,  20, 40,  60, 80},
                xtick={0,2,4,6,8,10,12,14,16,18,20},
                ylabel style ={font=\small},
                xlabel style ={font=\small},
                tick label style={font=\small}
                ]
                \addplot[draw=cg1, thick, densely dotted]        table[x=bitflip, y=BinaryDenseNet45]{\dataset};
                \addplot[draw=cg1, thick, densely dashed]       table[x=bitflip, y=BinaryDenseNet37]{\dataset};
                \addplot[draw=cg1, thick, dashdotted]               table[x=bitflip, y=BinaryDenseNet28]{\dataset};
                \addplot[draw=gray!80, thick]                            table[x=bitflip, y=BinaryResNetE18]{\dataset};
                \addplot[draw=gray!80, thick, densely dotted]   table[x=bitflip, y=RealToBinaryNet]{\dataset};
                \addplot[draw=gray!80, thick, densely dashed] table[x=bitflip, y=BinaryAlexNet]{\dataset};
                \addplot[draw=gray!80, thick, dashdotted]     table[x=bitflip, y=MeliusNet22]{\dataset};
                \addplot[draw=cg4, thick]                     table[x=bitflip, y=BiRealNet]{\dataset};
                \addplot[draw=red, thick]                     table[x=bitflip, y=XNORNet]{\dataset};
                \legend{BinaryDenseNet45, BinaryDenseNet37,
                        BinaryDenseNet28, BinaryResNetE18, RealToBinaryNet, BinaryAlexNet,
                        MeliusNet22, BiRealNet, XNORNet}
            \end{axis}
        \end{tikzpicture}
    }

    \vspace{-3mm}
    \subfloat[]{
        \begin{tikzpicture}
        % stuck-at zoo
            \pgfplotstableread[row sep=\\, col sep=&]{
            stuckat & BinaryDenseNet37Dilated & BinaryDenseNet45 & BinaryDenseNet37 & BinaryDenseNet28 & BinaryResNetE18 & RealToBinaryNet & BinaryAlexNet & MeliusNet22 & BiRealNet & XNORNet\\
            0.0 & 45.54 & 58.2 & 63.04 & 36.26 & 65.78 & 58.66 & 61.46 & 63.66 & 64.02 & 36.32 \\
            0.25 & 0.16 & 16.4 & 59.52 & 34.78 & 64.96 & 53.54 & 59.88 & 60.3 & 61.36 & 34.96 \\
            0.5 & 0.08 & 1.32 & 44.18 & 27.66 & 64.2 & 38.42 & 53.74 & 49.58 & 48.28 & 24.34 \\
            0.75 & 0.08 & 0.28 & 18.18 & 17.9 & 61.16 & 13.82 & 40.68 & 27.26 & 24.44 & 13.1 \\
            1.0 & 0.08 & 0.18 & 4.62 & 8.6 & 55.84 & 1.72 & 22.28 & 9.44 & 7.84 & 5.68 \\
            1.25 & 0.08 & 0.2 & 1.48 & 3.68 & 47.74 & 0.42 & 8.64 & 3.44 & 3.7 & 2.7 \\
            1.5 & 0.08 & 0.22 & 0.58 & 1.56 & 36.5 & 0.28 & 3.16 & 1.4 & 2.32 & 1.46 \\
            1.75 & 0.08 & 0.14 & 0.3 & 0.62 & 24.14 & 0.16 & 1.54 & 1.0 & 1.38 & 1.02 \\
            2.0 & 0.08 & 0.1 & 0.26 & 0.32 & 13.08 & 0.16 & 1.0 & 0.8 & 0.94 & 1.0\\
            }\dataset
            \begin{axis}[
                ylabel={Accuracy (\%)},
                xlabel={Injection rate (\%)},
               ylabel style={at={(-0.06, 0.5)}, anchor=north},
                legend style={at={(0.5,1.25)}, font=\tiny, anchor=north, legend cell align={left}},
                legend image post style={scale=0.7},
                legend columns=3,
                width=\textwidth,
                ymajorgrids = true,
                xmajorgrids = true,
                height=3.5cm,
                ymin=-5, ymax=70,
                xmin=-.05, xmax=2.01,
                     ytick={0,  20, 40,  60, 80},
                xtick={0, 0.25, 0.5, 0.75, 1.0, 1.25, 1.75, 2.0},
                ylabel style ={font=\small},
                xlabel style ={font=\small},
                tick label style={font=\small}
                ]
                \addplot[draw=cg1, thick, loosely dotted]     table[x=stuckat, y=BinaryDenseNet45]{\dataset};
                \addplot[draw=cg1, thick, densely dashed]     table[x=stuckat, y=BinaryDenseNet37]{\dataset};
                \addplot[draw=cg1, thick, dashdotted]         table[x=stuckat, y=BinaryDenseNet28]{\dataset};
                \addplot[draw=gray!80, thick]                 table[x=stuckat, y=BinaryResNetE18]{\dataset};
                \addplot[draw=gray!80, thick, loosely dotted] table[x=stuckat, y=RealToBinaryNet]{\dataset};
                \addplot[draw=gray!80, thick, densely dashed] table[x=stuckat, y=BinaryAlexNet]{\dataset};
                \addplot[draw=gray!80, thick, dashdotted]     table[x=stuckat, y=MeliusNet22]{\dataset};
                \addplot[draw=cg4, thick, dashdotted]         table[x=stuckat, y=BiRealNet]{\dataset};
                \addplot[draw=red, thick]                     table[x=stuckat, y=XNORNet]{\dataset};
            \end{axis}
        \end{tikzpicture}
    }

    \vspace{-3mm}
    \subfloat[]{
        \begin{tikzpicture}
        % dynamic zoo
            \pgfplotstableread[row sep=\\, col sep=&]{
            dynamic & BinaryDenseNet37Dilated & BinaryDenseNet45 & BinaryDenseNet37 & BinaryDenseNet28 & BinaryResNetE18 & RealToBinaryNet & BinaryAlexNet & MeliusNet22 & BiRealNet & XNORNet\\
            0.0 & 0.14 & 0.32 & 0.46 & 0.14 & 6.24 & 0.46 & 2.42 & 2.98 & 1.78 & 1.62 \\
            1.0 & 10.94 & 0.66 & 12.54 & 1.02 & 51.72 & 2.44 & 34.38 & 32.5 & 26.68 & 14.32 \\
            2.0 & 29.26 & 9.26 & 48.32 & 10.94 & 60.38 & 22.44 & 51.54 & 50.44 & 50.46 & 27.92 \\
            3.0 & 32.86 & 19.72 & 54.22 & 21.66 & 61.66 & 39.6 & 55.78 & 56.86 & 57.3 & 32.4 \\
            4.0 & 38.02 & 38.86 & 58.18 & 28.28 & 63.52 & 47.92 & 58.86 & 60.08 & 60.32 & 35.32 \\
            5.0 & 40.74 & 39.66 & 59.86 & 30.04 & 63.68 & 50.68 & 59.1 & 60.84 & 62.2 & 34.42\\
            }\dataset
            \begin{axis}[
                ylabel={Accuracy (\%)},
                xlabel={\# of XNOR ops},
               ylabel style={at={(-0.06, 0.5)}, anchor=north},
                legend style={at={(0.5,1.25)}, font=\tiny, anchor=north, legend cell align={left}},
                legend image post style={scale=0.7},
                legend columns=3,
                width=\textwidth,
                ymajorgrids=true,
                xmajorgrids=true,
                height=3.5cm,
                ymin=-5, ymax=70,
                xmin=-.1, xmax=5.1,
                    ytick={0,  20, 40,  60, 80},
                xtick={0, 1, 2, 3, 4, 5},
                ylabel style ={font=\small},
                xlabel style ={font=\small},
                tick label style={font=\small}
                ]
                \addplot[draw=cg1, thick, loosely dotted]                       table[x=dynamic, y=BinaryDenseNet45]{\dataset};
                \addplot[draw=cg1, thick, densely dashed]                       table[x=dynamic, y=BinaryDenseNet37]{\dataset};
                \addplot[draw=cg1, thick, dashdotted]                           table[x=dynamic, y=BinaryDenseNet28]{\dataset};
                \addplot[draw=gray!80, thick]                           table[x=dynamic, y=BinaryResNetE18]{\dataset};
                \addplot[draw=gray!80, thick, loosely dotted]                           table[x=dynamic, y=RealToBinaryNet]{\dataset};
                \addplot[draw=gray!80, thick, loosely dashed]                           table[x=dynamic, y=BinaryAlexNet]{\dataset};
                \addplot[draw=gray!80, thick, dashdotted]                           table[x=dynamic, y=MeliusNet22]{\dataset};
                \addplot[draw=cg4, thick, dashdotted]                           table[x=dynamic, y=BiRealNet]{\dataset};
                \addplot[draw=red, thick,mark options={fill=red}] table[x=dynamic, y=XNORNet]{\dataset};
            \end{axis}
        \end{tikzpicture}
    }
    \vspace{-1mm}
    \caption{Simulation results of (a) bit-flips, (b) stuck-at, and (c) dynamic faults on different models.}
    \label{fig:fault_models}
\vspace{-2mm}
\end{figure*}
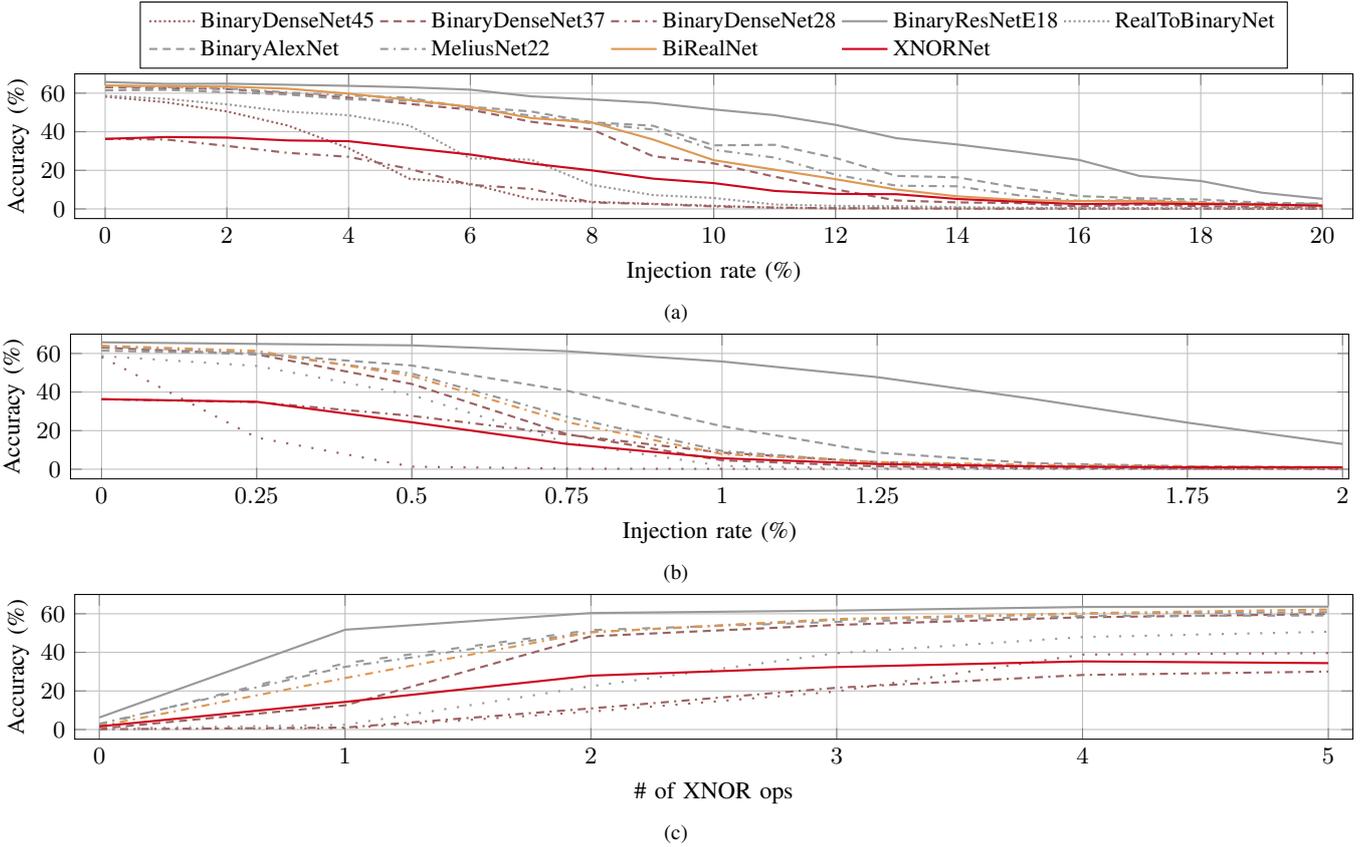

This section discusses the simulation results. Table~\ref{tab:sim_spec} shows the system specifications used to conduct all experiments. We verified the functionality of FLIM in two distinct experiments. The fault injector extends the Tensorflow/Larq framework. Hence, we compared the inference results of FLIM (without injecting any faults) with the results of vanilla Tensorflow/Larq. The fault distribution and mapping have been verified with X-Fault. Our investigations exhibit the impact of faults on BNNs from various perspectives. First, the impact on individual layers is studied. Second, we compare the performance of our simulator to X-Fault and vanilla Tensorflow. Finally, we thoroughly explore the resilience of various models on bit-flip and stuck-at faults.

\begin{table}[ht]
    \centering
    \caption{Adopted experimental setup.}
    \resizebox{0.6\columnwidth}{!}{%
        \begin{tabular}{l|l}
            \multicolumn{2}{l}{\textbf{Hardware}} \bigstrut[b]\\
            \hline
            CPU & AMD Ryzen 7 5800X \bigstrut\\
            \hline
            RAM & DDR4 2666MHz 64GB \bigstrut\\
            \hline
            GPU & NVIDIA GeForce RTX 3080 Ti 12GB \bigstrut[t]\\
            \hline
            \hline
            \multicolumn{2}{l}{\textbf{Software}} \bigstrut[b]\\
            \hline
            GPU Driver & 470.129.06 \bigstrut\\
            \hline
            CUDA & 11.4 \bigstrut\\
            \hline
            CuDNN & 8.1.0.77-1 \bigstrut\\
            \hline
            TensorFlow & 2.8.0 \bigstrut\\
            \hline
            LARQ & 0.12.0 (modified) \bigstrut[t]\\
        \end{tabular}%
    }
    \label{tab:sim_spec}%
\end{table}%

\textbf{Layer resilience:} This experiment aims to investigate the resilience of individual layers of a BNN. We use a binary version of LeNet~\cite{LeCun1989} trained on the MNIST dataset. LeNet represents a convolutional neural network which, in this experiment, consists of three convolutional layers and two dense layers. The former aims to extract the visual features from the input picture. The latter is responsible for the feature classification. The MNIST dataset embodies a set of 28$\times$28 greyscale pixel images depicting handwritten digits~\cite{LeCun1998}. After training, the model achieves an accuracy of $97.62\%$ without any injected faults.

Throughout the experiment, each layer is mapped onto a single crossbar while sweeping the injection rate of bit-flips, dynamic faults, and stuck-at faults. To mitigate the impact of randomly placing the faults on the crossbar, we performed every experiment hundred times which reinitialized the random generator with a new seed value.

Fig.~\ref{fig:fault_layers}(a-b) illustrates that stuck-at faults impact the model more severely than bit-flips independent of the layer. \textit{While stuck-at faults influence almost all layers equally strongly, bit-flip faults affect the accuracy depending on the layer depth. Moreover, convolutional layers appear more susceptible to bit-flips than dense layers}. The impact of dynamic bit-flip faults is shown in Fig.~\ref{fig:fault_layers}(c), whereas the x-axis represents the number of XNOR operations required to sensitize the fault. The results show that the BNN model's accuracy stabilizes around its original value at around four consecutive XNOR operations.

Next, we investigate the impact of faulty rows/columns on the model's accuracy. This experiment instantiates a 40$\times$10 crossbar for each layer. Fig.~\ref{fig:fault_layers}(d-e) portrays the results of this experiment. Once again, the layer's depth directly correlates with the impact on accuracy. In particular, the last dense layer declines almost linearly. In general, the impact of faulty columns is more substantial than of faulty rows. Considering the column-wise parallelism of XNOR operations, this result appears plausible.

\textbf{Performance evaluation:} We evaluate the performance of our fault injection platform by executing the inference on the previous LeNet model together with the complete MNIST test dataset consisting of 10.000 images. While FLIM and the vanilla Larq implementation perform fifty consecutive runs of the complete dataset, we estimate the total run time of X-Fault based on five images. During the inference, the fault injection mechanism maps the respective operations but does not inject actual faults. Thus, the vanilla Larq implementation serves as a lower boundary regarding the total simulation time.

Fig.~\ref{fig:fault_layers}(f) shows the substantial performance improvement of our work. \textit{FLIM classifies the 10.000 images 29375$\times$ faster than X-Fault.} Due to the deep integration within Larq and Tensorflow, FLIM takes advantage of GPUs, \textit{doubling the performance to a speed-up of 66754$\times$ compared to X-Fault.} Conclusively, FLIM abstracts the fault model on the XNOR operation level and, hence, trades simulation accuracy with noteworthy performance improvement.

\textbf{Model resilience:} The last experiment investigates the resilience of various models (see Table~\ref{tab:model}). We pre-trained the models with the ImageNet~\cite{Russakovsky2015} dataset and injected bit-flips, dynamic, and stuck-at faults. Once again, we run every experiment hundred times to mitigate the impact of the randomly placed faults.

Fig.~\ref{fig:fault_models}(a-c) displays the simulation results. As expected, the obtained results indicate that stuck-at faults cause a more substantial impact on the accuracy than bit-flips. \textit{In other words, it is possible to see that faults related to time-dependent deviations can affect the reliability of emerging applications differently. Depending on the injection rate, transient faults will compromise the reliability of such applications at different levels. In addition, it is possible to see that the reliability of emerging applications is more affected by permanent faults.} The BiRealNet and XNOR-Net represent a particular case because their convolutions are not strictly binarized. BiRealNet utilizes real-valued activation functions through identity shortcuts~\cite{Liu2020}. On the other hand, XNOR-Net's weights are multiplied by an individual gain based on the magnitude of the channel. Still, FLIM is capable of simulating both models by slightly adjusting the bit-flip mask.

\begin{table}[ht]
    \centering
    \caption{Overview of the BNN models and their characteristics.}
    \resizebox{\columnwidth}{!}{%
        \begin{tabular}{l|c|c|c|c|c}
            \textbf{Model} & \multicolumn{1}{p{5em}|}{\textbf{Top-1 Acc. }} & {\textbf{Size}} & \textbf{Parameters} & \textbf{MACs} & \textbf{Binarized} \bigstrut[b]\\
            \hline
            RealToBinaryNet~\cite{Martinez2020} & 65.0\% & 5.13MB & 12M & 1.81B & 92.39\% \bigstrut\\
            \hline
            BinaryDenseNet45~\cite{Bethge2019} & 65.0\% & 7.54MB & 13.9M & 6.67B & 96.34\% \bigstrut\\
            \hline
            BinaryDenseNet37~\cite{Bethge2019} & 62.9\% & 5.25MB & 8.7M & 4.71B & 96.76\% \bigstrut\\
            \hline
            BinaryDenseNet28~\cite{Bethge2019} & 60.9\% & 4.12MB & 5.13M & 3.79B & 94.66\% \bigstrut\\
            \hline
            BinaryResNetE18~\cite{He2016} & 58.3\% & 4.03MB & 11.7M & 1.81B & 92.4\% \bigstrut\\
            \hline
            BinaryAlexNet~\cite{Krizhevsky2017} & 36.3\% & 7.49MB & 61.8M & 841M & 91.34\% \bigstrut\\
            \hline
            MeliusNet22Z\cite{Bethge2021} & 62.9\% & 3.88MB & 6.94M & 4.76B & 97.14\% \bigstrut\\
            \hline
            Bi-Real Net~\cite{Liu2020} & 57.5\% & 4.03MB & 11.7M & 1.81B & 92.4\% \bigstrut\\
            \hline
            XNORNet~\cite{Rastegari2016} & 45.0\% & 22.81MB & 62.4M & 1.14B & 90.05\% \bigstrut[t]\\
        \end{tabular}%
    }
    \label{tab:model}%
\end{table}%

%\vspace{-2mm}
\section{Conclusion}
\label{sec:con}
This work proposed a fault injection platform, called \textbf{FLIM}, able to evaluate the impact of in-field faults related to time-dependent deviations in emerging applications. The platform injects bit-flips (static and dynamic), related to environmental variations, and stuck-at faults, associated with temporal variations. We investigated the impact of these faults on individual layers and various models. Furthermore, FLIM outperforms the current state-of-the-art platform by four orders of magnitude in terms of performance. The obtained results show that a certain level of in-field faults can be tolerated and that the impact of bit-flips, even if multiple, compromises the reliability of emerging applications less than stuck-at faults. These results also demonstrate that in order to guarantee the development of high-reliability emerging applications, it is mandatory to adopt not only fault-tolerant approaches but also strategies able to monitor and/or mitigate applications' degradation during their lifetime. In the future, we want to extend the capabilities of FLIM to inject faults during training.

%\vspace{-4mm}
\balance
\bibliographystyle{IEEEtran}
\bibliography{bibliography.bib}
\end{document}